\documentclass[namedreferences]{solarphysics}
\usepackage[optionalrh,solaromanenum]{spr-sola-addons} 
\usepackage{graphicx}                    
\usepackage{color}                       
\usepackage{url}                         

\begin{document}

\begin{article}

\begin{opening}

\title{Magnetography of Solar Flaring Loops with Microwave Imaging Spectropolarimetry}

%
\author{D. E.~\surname{Gary}$^{1}$\sep
        G. D.~\surname{Fleishman}$^{1}$\sep
        G. M.~\surname{Nita}$^{1}$
       }

%
\runningauthor{Gary et al.}
\runningtitle{Magnetography of Flaring Loops}

%
  \institute{$^{1}$ Physics Dept., New Jersey Institute of Technology
                     email: \url{dgary@njit.edu} email: \url{gfleishm@njit.edu} email: \url{gnita@njit.edu}
             }

\begin{abstract}
We have developed a general framework for modeling gyrosynchrotron and free-free emission from solar flaring loops and used it to test the premise that 2D maps of source parameters, particularly magnetic field, can be deduced from spatially resolved microwave spectropolarimetry data.  In this paper we show quantitative results for a flaring loop with a realistic magnetic geometry, derived from a magnetic field extrapolation, and containing an electron distribution with typical thermal and nonthermal parameters, after folding through the instrumental profile of a realistic interferometric array.  We compare the parameters generated from forward fitting a homogeneous source model to each line of sight through the folded image data cube with both the original parameters used in the model and with parameters generated from forward fitting a homogeneous source model to the original (unfolded) image data cube.  We find excellent agreement in general, but with systematic effects that can be understood as due to finite resolution in the folded images and the variation of parameters along the line of sight, which are ignored in the homogeneous source model.  We discuss the use of such 2D parameter maps within a larger framework of 3D modeling, and the prospects for applying these methods to data from a new generation of multifrequency radio arrays now or soon to be available.

\end{abstract}

%

\end{opening}

%

\section{Introduction}


The magnetic structure of the solar corona plays a key role in all of solar activity. For example, in a recent dedicated review, \inlinecite{Aschwanden_2008} identifies ten outstanding problems in Solar Physics. Five of them---hydrodynamics of coronal loops, MHD oscillations and waves (coronal seismology), coronal heating, magnetic reconnection, and particle acceleration---require measurement of coronal parameters, especially the magnetic field, \textit{in or near the flaring region} and on dynamical time scales.

However, direct measurements of the magnetic field in the tenuous atmosphere are extremely difficult to make.  Instead, the field strength and direction are measured at non-flaring times at the photospheric (or possibly chromospheric) boundary; specifically, vector fields are measured from full-Stokes polarized intensity of Zeeman sensitive spectral lines with circular polarization giving line of sight field strength and linear polarization providing the transverse field. Then, to assess the coronal magnetic field these measured photospheric fields are extended into the corona through potential or force-free field extrapolations.  However, even extrapolations with excellent data can yield incorrect results \cite{DeRosa_etal_2009} due to several limitations: (1) the photosphere does not meet the force-free condition on which the extrapolations are generally based, (2) the curved boundary of the solar surface and resulting near-limb foreshortening complicates the geometry of the extrapolation, (3) the line profiles on which the measurements are based can be affected by non-LTE and Doppler effects, and (4) the measurements are affected by the 180-degree ambiguity (only the angle, not direction of the transverse field is measured), scattered light, and evolution of the region during the measurements.  Some of these limitations can be addressed.  For example, measurements at the more force-free chromospheric boundary are possible (e.g. \opencite{Socas-Navarro_etal_2006}), vector photospheric measurements can be preprocessed, i.e., modified to approximate the force-free boundary condition, with reference to other observations such as H$\alpha$ fibrils, within observational errors of the transverse photospheric field measurements \cite{Wiegelmann_etal_2006, Wiegelmann_etal_2008},  and techniques exist to resolve the 180-degree ambiguity  (e.g. \opencite{Metcalf_etal_2006}).  But even so, such methods give rise to modeled, not measured, pre-event coronal magnetic fields, which cannot follow the relevant dynamical changes that occur in flares.

More indirect observational clues to the \textit{coronal magnetic field} can sometimes be exploited, such as morphological tests at the level of the chromosphere (H$_\alpha$ fibrils aligned with magnetic field direction, e.g. \opencite{Wiegelmann_etal_2008}) and corona (EUV and soft X-ray loops, although these provide only the shape of field lines, and even this has been called into question by the work of \opencite{Mok_etal_2008}, who showed through 3D modeling that apparent loops based on brightness do not necessarily reveal the underlying field line shapes).  Direct coronal magnetic field measurements through Zeeman splitting of infrared lines has been attempted \cite{Lin_etal_2000,Lin_etal_2004,Liu_Lin_2008}, and more are planned \cite{Tomczyk_2012}, but they require a long accumulation time and apply only along extended lines of sight above the solar limb where the extrapolations are most difficult, frustrating attempts at direct comparison of observations and models. To properly address the outstanding theoretical problems cited by \inlinecite{Aschwanden_2008} it is essential to seek additional, independent techniques for directly measuring the coronal field, especially the dynamically changing fields in the flaring region.

Diagnostics of \textit{coronal thermal structure} are obtainable with a combination of EUV filtergram images and spectral line measurements, although such issues as atomic and ionic species abundances, multithermal plasma along the line of sight, and non-LTE effects make interpretation ambiguous and difficult.  The emission-measure- (density-squared-) weighted brightness also makes some regions of tenuous plasma too faint for such diagnostic information, while any flaring regions tend to saturate the detectors. Information on the high-energy component of the flaring plasma requires soft- and hard-X-ray observations, thus requiring the piecing together of information from multiple spacecraft, pertaining to different regions of the corona and generally available for only a subset of events.

In this paper we outline a practical method of coronal diagnostics that in principle can achieve the required dynamical measurement of coronal magnetic field, thermal plasma, and particles: broadband microwave imaging spectropolarimetry, augmented by sophisticated modeling and forward fitting. The feasibility of this approach has already been proven by numerical tests \cite{Bastian_etal_1998,Bastian_2006, Fl_etal_2009}, assuming a hypothetical, ideal radio heliograph providing data with arbitrarily high spatial resolution. The forward fitting of actual radio spectra has been attempted with some success for a limited number of events, whose spatially-integrated microwave emission (recorded in the form of total power spectra) could be modeled by a relatively uniform source or a combination of two uniform sources \cite{Bastian_etal_2007,Altyntsev_etal_2008,Fl_etal_2011,Fl_etal_2013}. However, in the more common case of an inhomogeneous flaring region, the quantitative diagnostics of flaring loops requires imaging spectroscopy and polarimetry data, which have not yet been routinely available. Fortunately, the recent advances in radio interferometric imaging instruments (Janksy Very Large Array---JVLA, Expanded Owens Valley Solar Array---EOVSA, and others) will soon provide for the first time the level of microwave imaging spectropolarimetry and unprecedented data quality necessary to deduce the key thermal and non-thermal plasma parameters, including magnetic field strength, needed for detailed coronal diagnostics.  It is therefore timely to explore the potential for such soon-to-be-realized instruments to exploit microwave imaging spectropolarimetry for practical measurement of dynamically changing coronal plasma parameters.

In \S~\ref{model_framework} we describe the methodology we use to create a realistic model flaring loop and its multifrequency polarized radio emission for further study, using the EOVSA instrument profile as a specific example for comparison purposes.  In \S~\ref{fitting_framework} we describe the method of forward-fitting theoretically determined microwave spectra to each resolution element of the model to obtain 2D parameter maps.  In \S~\ref{results} we describe quantitative comparisons of the fitted parameters with the model, to show that even with the finite spatial resolution and image quality expected from EOVSA it is possible to obtain both quantitative and qualitative information on the dynamically changing plasma parameters, including magnetic field strength and direction along the flaring loop.  We conclude in \S~\ref{conclusions}, and provide an outline of a more ambitious 3D modeling framework that could be developed in the future to overcome the unavoidable distortions caused by finite resolution.

\section{Modeling Framework \label{model_framework}}
\begin{figure}    
\centerline{\includegraphics[width=0.9\textwidth,clip=]{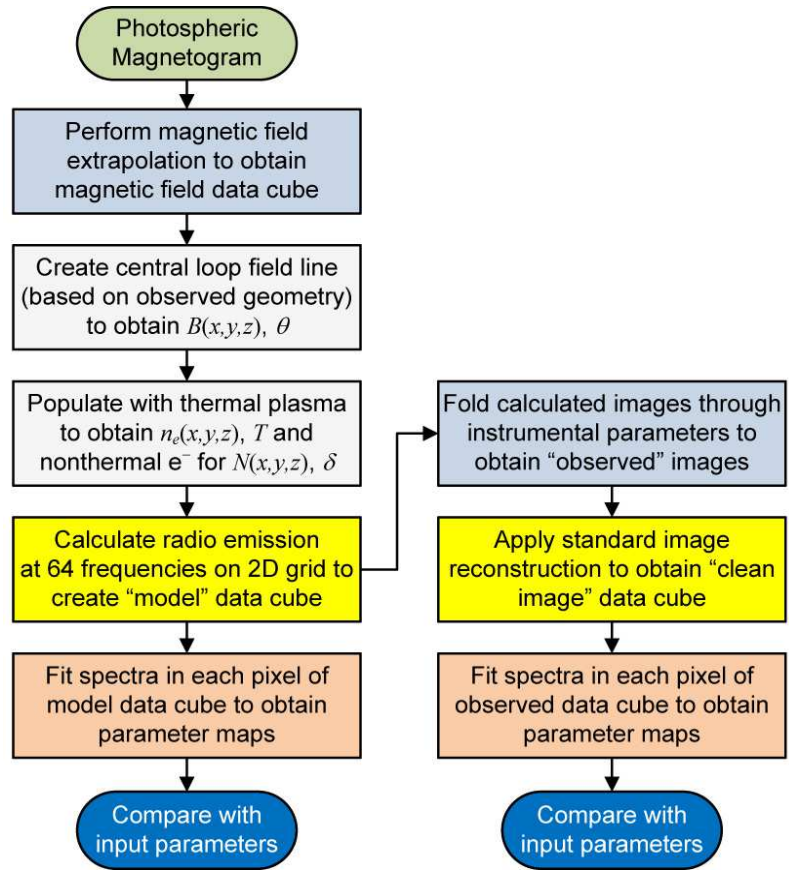}}
\caption{Block diagram showing the steps performed to create a flaring loop model, use it to calculate multifrequency radio emission from the model (the unfolded data cube), and fold the images of the data cube through an actual instrument to obtain the folded data cube. }
\label{Alg_1}
\end{figure}

We give in Figure~\ref{Alg_1} a block diagram showing the steps in the modeling framework that will be described in this section.  There are three main steps: (1) create a model (first three blocks on the left in Fig.~\ref{Alg_1}), (2) calculate the multifrequency radio emission from the model (yellow block on the left in Fig.~\ref{Alg_1}), and (3) fit the calculated radio spectra (bottom, pink block on the left in Fig.~\ref{Alg_1}).  However, to include the effects of finite resolution and noise introduced by an actual instrument, we also include the steps in the right column of Fig.~\ref{Alg_1}.  We now describe each of these steps in turn.

\subsection{Method of Generation of Flaring Loop Model}
To generate the spatial geometry of the flaring loop, we start with a tool that we have developed called the GX\_simulator\footnote{GX Simulator is now a part of the Solarsoft (SSW) IDL distribution.} \cite{Nita_etal_2011,Nita_etal_2012}, which provides a graphical interface for creating and manipulating magnetic field models including those generated from extrapolation of photospheric magnetograms.  In the case used for illustration in this paper, we started with a nonlinear force-free (NLFF) magnetic field extrapolation of a Hinode vector magnetogram embedded in a wider-field MDI magnetogram for active region 10956, taken on 2007 May 18, although we note that our results are not highly dependent on the choice of magnetic field model.  We seek only a loop with a realistic geometry, in this case one with a broad range of magnetic field up to reasonably high magnetic field strength, a moderately high mirror ratio, and asymmetric magnetic footpoints.  Using the tool, we identify a suitable magnetic field line to use as the central field line of the loop (footpoint fields $B_1 = 1090$ G, $B_2 =664$ G; mirror ratio $B_1/B_{\rm min} = 9.1$), loop length $l = 6.478\times10^9$ cm, and then impose the thermal plasma and an energetic electron distribution with parameterized spatial extent both along ($s$) and across ($x$, $y$) the loop, wider at the apex and tapering at the footpoints in accordance with the conservation of magnetic flux.  We parameterize the thermal electron distribution with a temperature $T = 2\times10^7$ K; thermal density that is hydrostatic in height $z$
\begin{equation}
n_e = n_0 \exp\Bigl[{-\left({x\over a}\right)^2 - \left({y\over b}\right)^2}\Bigr] \exp\Bigl[{-{z/R\over6.76\times10^{-8}T}}\Bigr],
\end{equation}
where $n_0 = 5\times10^9$ cm$^{-3}$ is the on-axis base density and $a = b = 4.37$~Mm are the off-axis scale distances at the loop apex (point where $B=B_{\rm min}$), perpendicular to the loop, scaling at other points on the axis as $B_{\min}/B$; number density of nonthermal electrons \begin{equation}
N_e = n_b \exp\Bigl[-\left({x\over a}\right)^2 -\left({y\over b}\right)^2\Bigr] \exp\left[-2\left({s - s_0 \over l}\right)^2\right],
\end{equation}
where $n_b = 3\times10^6$ cm$^{-3}$, $a$ and $b$ are as above, and $s_0 = -0.287\;l$ is the location of the peak of the nonthermal density relative to the loop apex, negative in the direction toward $B_1$; powerlaw in energy $N_e = (\delta-1)(n_b/E_0)(E/E_0)^{-\delta}$ with index $\delta = 5$; and high- and low-energy cutoffs $E_{\rm max}=10$~MeV and $E_{\rm min}=0.1$~MeV, respectively.  For this model, we use an isotropic pitch-angle distribution, and the only electron parameters that vary spatially are $n_e$ and $N_e$.

We emphasize that for this model it is not essential that we have a physically consistent set of parameters---we seek only a model with sufficient complexity to be a good test of parameter recovery via spatially resolved microwave spectropolarimetry.  The spatial complexity comes from the spatial variation of $B$, $\theta$ (angle between the line of sight and $B$), $n_e$, and $N_e$.  In principle we could also have allowed the electron energy parameters $T$, $\delta$, $E_{\rm max}$, $E_{\rm min}$ and pitch angle to vary spatially also, but although such variation would be interesting to study (and indeed is planned for future work) it is not deemed essential to the point of this paper.

\begin{figure}    
\centerline{\includegraphics[width=0.9\textwidth,clip=]{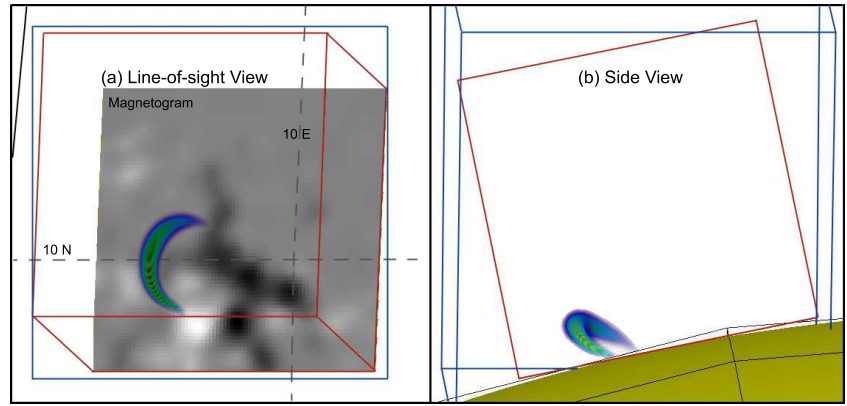}}

\caption{($a$) Line-of-sight view of the model loop, showing the base magnetogram as the gray-scale image and the distribution of nonthermal electrons. ($b$) Same as in $a$, but shown as a side-view aligned with the line-of-sight. The solar surface is shown in yellow at the base of the loop.}

\label{Model_View}
\end{figure}

Once the loop and electron geometries are set, the loop may be oriented for any line of sight prior to calculating the emission.  For this test we oriented the loop near the center of the disk (heliographic coordinates E12, N11), but chose a central field line with a considerable angle to the line of sight, so that the variation of magnetic field direction along the loop ranges from $ 64^\circ < \theta < 123^\circ$.  Figure~\ref{Model_View} shows the final geometry of the loop.

\subsection{Generation of Microwave Emission from the Model}
\begin{figure}    
\centerline{\includegraphics[width=0.5\textwidth,clip=]{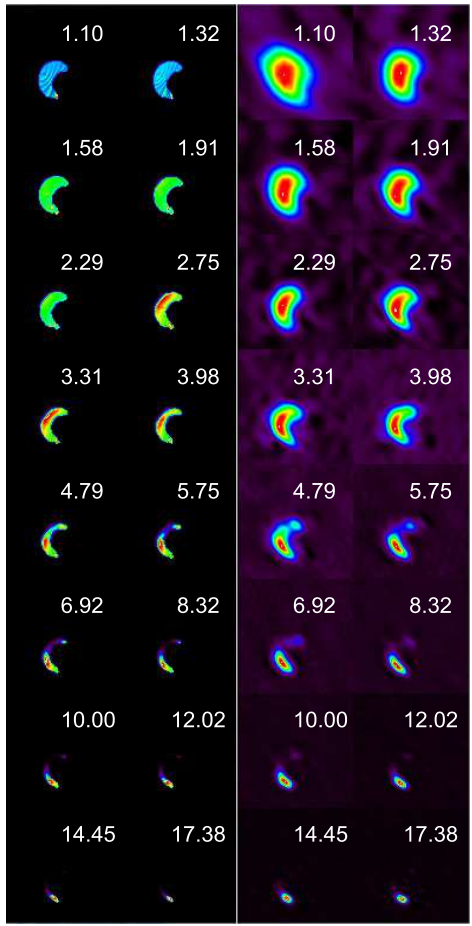}}
\caption{($a$) LCP images from the model at 16 selected frequencies distributed logarithmically from 1-18 GHz.  The number in each panel gives the frequency for that image in GHz.  The stripes in the lower-frequency images are predicted by the theory, and are due to brightness variations associated with the discrete, lower harmonics of the gyrofrequency. ($b$) Same as in $a$, but now folded through the EOVSA instrument and reconstructed using the CLEAN algorithm.}
\label{Data_Cube}
\end{figure}

Once the parameters of the flaring loop are set and the line-of-sight orientation is chosen, we calculate the microwave emission using the hybrid fast codes of \inlinecite{Fl_Kuzn_2010}. 
The emission from 1-18 GHz is calculated for each line of sight, with spatial (3D voxel) resolution of $2''$ on a side, at 64 logarithmically spaced frequencies in two polarizations. The hybrid codes use exact calculations for the low harmonics of the gyrofrequency, and highly accurate but approximate calculations for the high harmonics, separately in the two magnetoionic modes, from which the two senses of circular polarization can be derived.  The calculation is done considering each voxel as a homogeneous source using the parameters computed at the center of the voxel, but the resulting brightness of a 2D pixel takes into account the correct radiative transfer including frequency dependent mode coupling \cite{White_ea_1992,Bastian_1995} through all voxels along the line of sight.  These calculations are done directly from within the GX\_simulator tool, to generate the entire set of simulated multifrequency images (data cube) in just a few minutes.  Figure~\ref{Data_Cube}$a$ shows a 16-image sampling of the 64-image data cube in left circular polarization (LCP), although the calculations are actually done in both right circular and left circular polarizations (RCP and LCP, respectively).

\subsection{Folding Emission through the EOVSA Instrument Profile}
\begin{figure}    
\centerline{\includegraphics[width=0.9\textwidth,clip=]{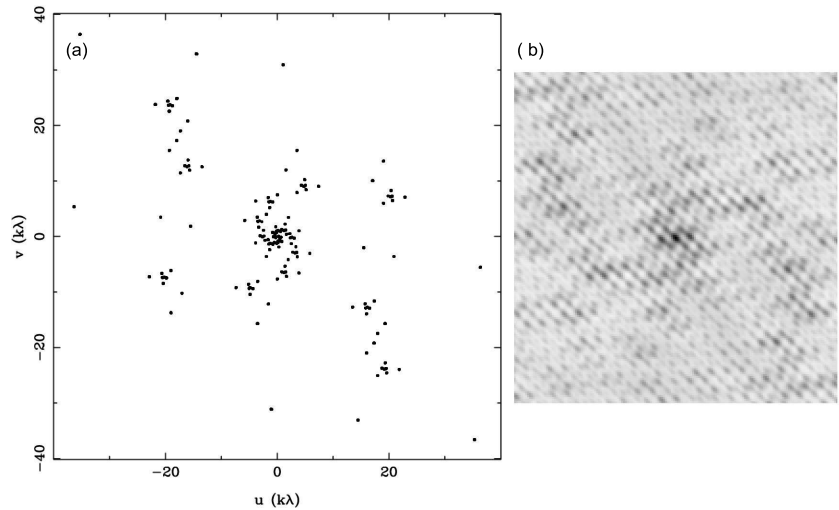}}
\caption{($a$) Sampling in the Fourier (uv) plane used for the simulation.  This represents the 78 baselines of EOVSA, at a frequency of 10~GHz, for a time near noon when the Sun is at +15$^\circ$ declination. ($b$) The point-spread-function (synthesized beam) corresponding to the uv coverage in $a$.  The gray scale is inverted for clarity.}
\label{UV_Beam}
\end{figure}

Since our goal is to assess the possibility of recovering flaring loop parameters from real data, we must choose a specific instrument through which to fold the input data generated from the model.  For this we select the Expanded Owens Valley Solar Array (EOVSA, \opencite{Gary_ea_2011}), which is a 13-antenna array currently under construction for completion in 2013. Figure~\ref{UV_Beam} shows the ``snapshot'' uv coverage and corresponding point-spread-function (synthesized beam) of EOVSA at a frequency of 10~GHz, near noon on a date when the Sun is at a declination of +15$^\circ$.

Using the Miriad radio interferometric imaging package \cite{Sault_ea_2011}, it is straightforward to use the individual images from the data cube as a brightness model, Fourier transform the brightness model to obtain the visibilities in the uv plane, and then sample those visibilities with the uv-coverage of Fig.~\ref{UV_Beam} to obtain a model-visibilities data set for each frequency and polarization.  To those visibilities are added a realistic level of random noise representing the thermal noise of the system (4500 K), and the standard CLEAN algorithm is then used to generate reconstructed images representing the images EOVSA would actually produce.  Figure~\ref{Data_Cube}$b$ shows the reconstructed LCP images for the same 16 frequencies as in Fig.~\ref{Data_Cube}$a$, where now the effect of the finite, frequency-dependent resolution is apparent.  From the original (unfolded) 64-image data cube we thus construct a parallel 64-image folded data cube representing the EOVSA reconstructed images in each circular polarization.

\subsection{Comparing the Unfolded Model with the Folded Model}
\begin{figure}    
\centerline{\includegraphics[width=0.5\textwidth,clip=]{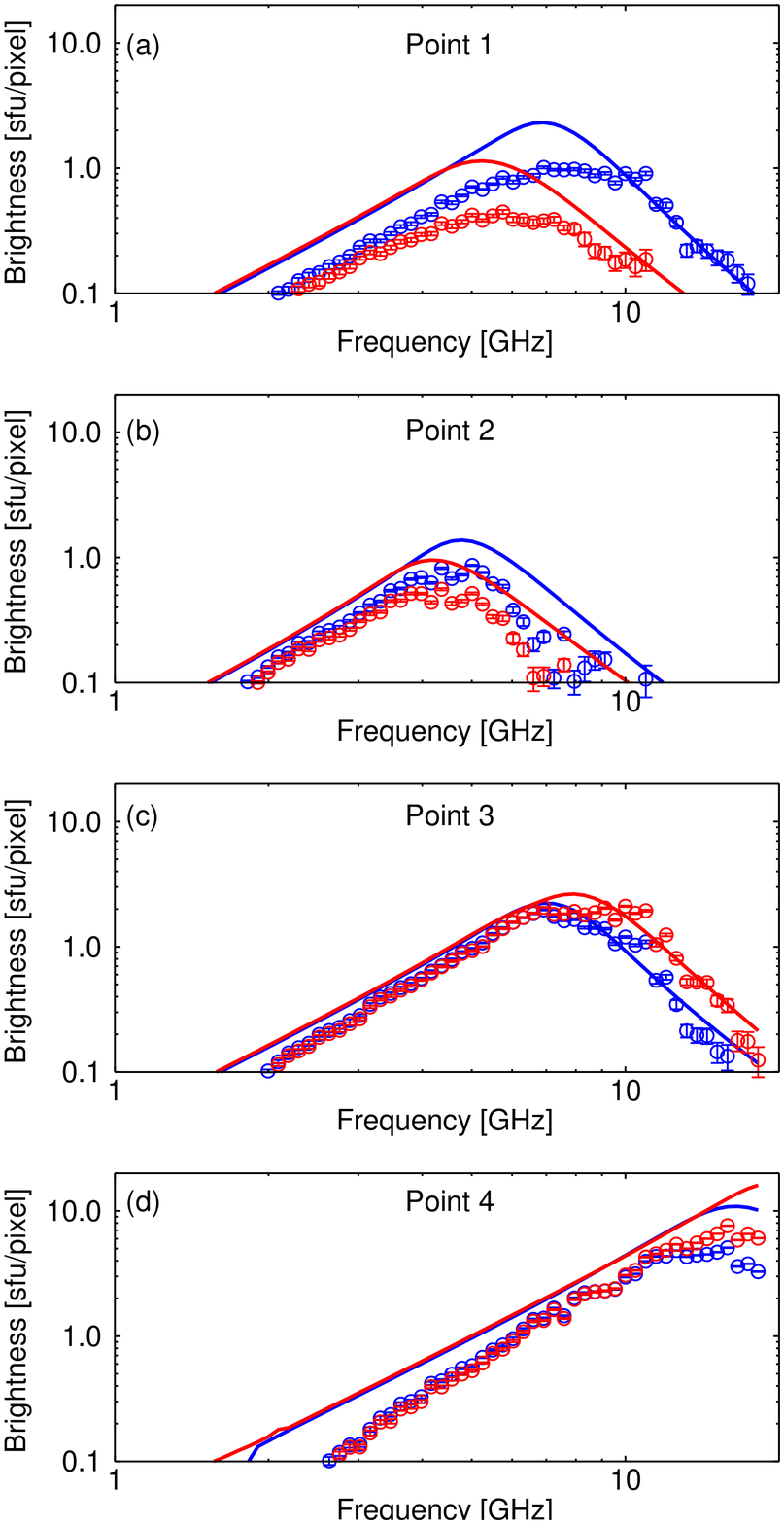}
            \includegraphics[width=0.4\textwidth,clip=,
            bb=150 -200 450 300]{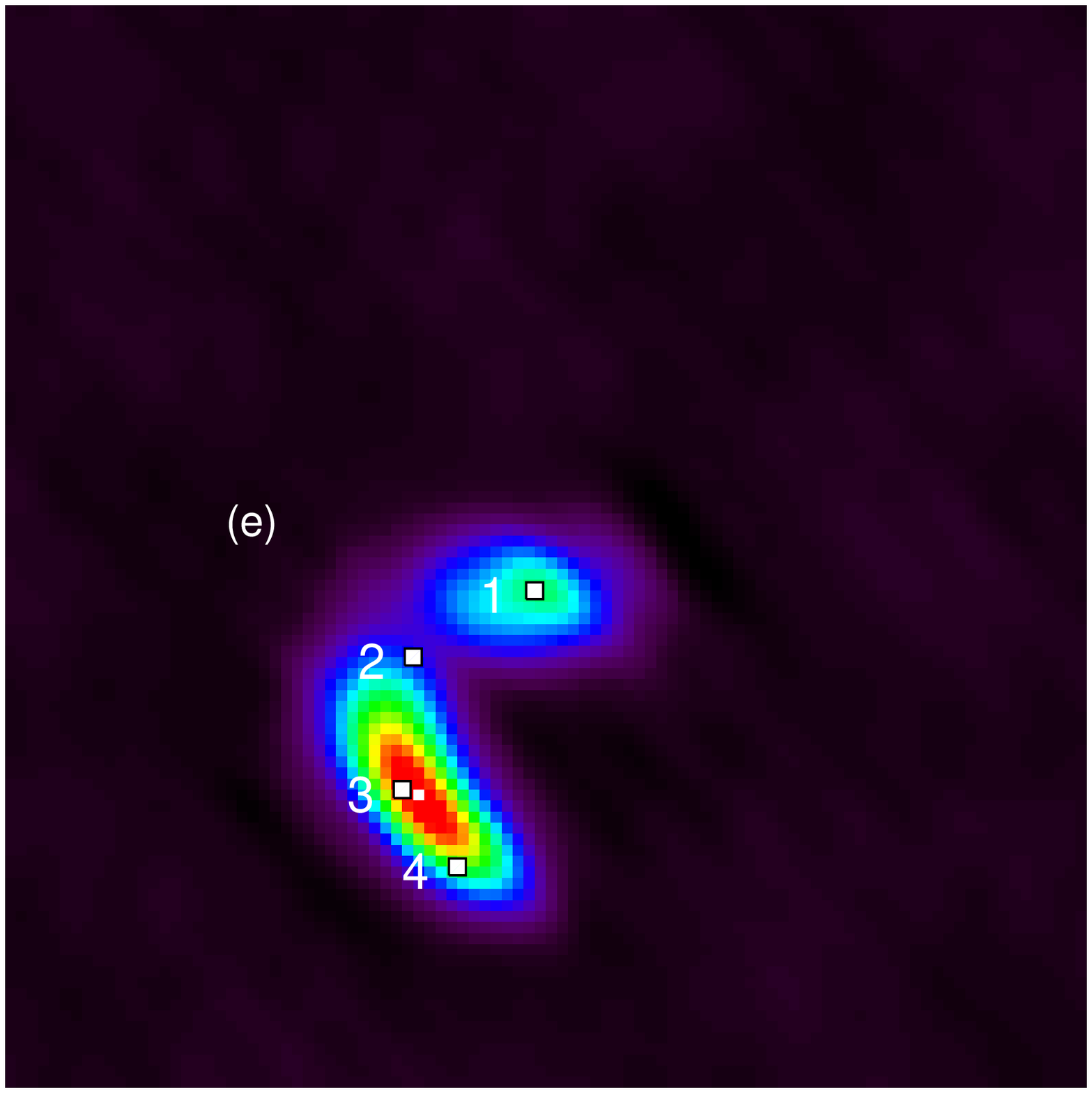} }
\caption{($a$-$d$) Comparison of raw spectra for four selected pixels in both the unfolded (lines) and folded (symbols) data cubes.  Red corresponds to RCP, while blue is LCP. ($e$) Folded RCP image at 6.3 GHz, showing the locations of the four points whose spectra are plotted in $a$-$d$.}
\label{Raw_spec}
\end{figure}

An alternative to viewing the datacube as a set of multifrequency images is to consider the third (spectral) dimension along various lines of sight, which provides spatially resolved, polarized brightness temperature spectra. Figure~\ref{Raw_spec} shows a comparison of polarized brightness temperature spectra sampled at a few locations in the loop to show how the finite resolution and image reconstruction affect the spectra.  In general, the folded spectra quite faithfully agree with the unfolded spectra, although there are systematic differences such as the lower flux density at low frequencies (due to the finite spatial resolution) and the corresponding steeper slope. This steeper slope also subtly shifts the peak of the spectrum of the folded data cube to slightly higher frequencies.  In addition, finite dynamic range causes the weaker emission at high frequencies (i.e. that near point 2, whose spectrum is shown in Fig.~\ref{Raw_spec}$b$) to be poorly reconstructed in the presence of bright emission in the same image.  The dynamic range of the images shown in Fig.~\ref{Data_Cube}$b$ ranges from 100:1 to 200:1.  Various strategies exist for improving the dynamic range of reconstructed images, such as the use of frequency synthesis \cite{Conway_ea_1990,Sault_Wieringa_1994,Rau_Cornwell_2011}, but they have not been used in this study.

From Fig.~\ref{Raw_spec}$a$-$d$ it can be seen that the spectral differences from pixel to pixel in the images, especially shifts in the peak frequency, are greater than the differences between unfolded and folded spectra for a given pixel.  This suggests that the spectral shape of the folded spectra are dominated by changes in physical parameters rather than by the effects of finite spatial resolution, and hence spectral fits to the folded spectra promise to provide reasonable estimates of source parameters.  We now proceed to demonstrate this by fitting both unfolded and folded spectra at each pixel in the two data cubes and comparing the fitted parameters.

\section{Fitting Framework\label{fitting_framework}}
Once the two parallel 2D data cubes (unfolded and folded) are available, we step through the 2D field of view pixel by pixel and apply forward fitting of a homogeneous source to the spectra via the downhill simplex minimization algorithm \cite{Press_etal_1986}, with some modifications detailed in \inlinecite{Fl_etal_2009} to avoid local minima.  The method is to choose a set of $N_{\rm par}$ parameters to fit, make an initial guess for parameter values, calculate the emission for that set of parameters using the fast codes, compare the fit with the data using a standard reduced-chi-squared ($\chi^2_\nu$) metric, adjust the parameters via the simplex method and repeat until some specified criterion for stopping has been met.  As described in  \inlinecite{Fl_etal_2009}, once a minimum is found the solution is ``shaken'' by strongly perturbing the solution vector, and the algorithm is repeated to perhaps find the same or a different minimum solution.  This is done until either the same minimum has been repeatedly found or the number of shakes is $N_{\rm par} + 6$.  At the end of this procedure, the solution with the smallest $\chi^2_\nu$ is taken.  For the results in this paper, the fitting was done using $N_{\rm par} = 6$, i.e. $B$, $\theta$, $n_e$, $N_e$, $\delta$ and $E_{\rm min}$.  The value of $E_{\rm max}$ is not well constrained for emission to 18~GHz, and was kept fixed at 5~MeV (a factor of two lower than the value actually used in the model).

\begin{figure}    
\centerline{\includegraphics[width=0.5\textwidth,clip=]{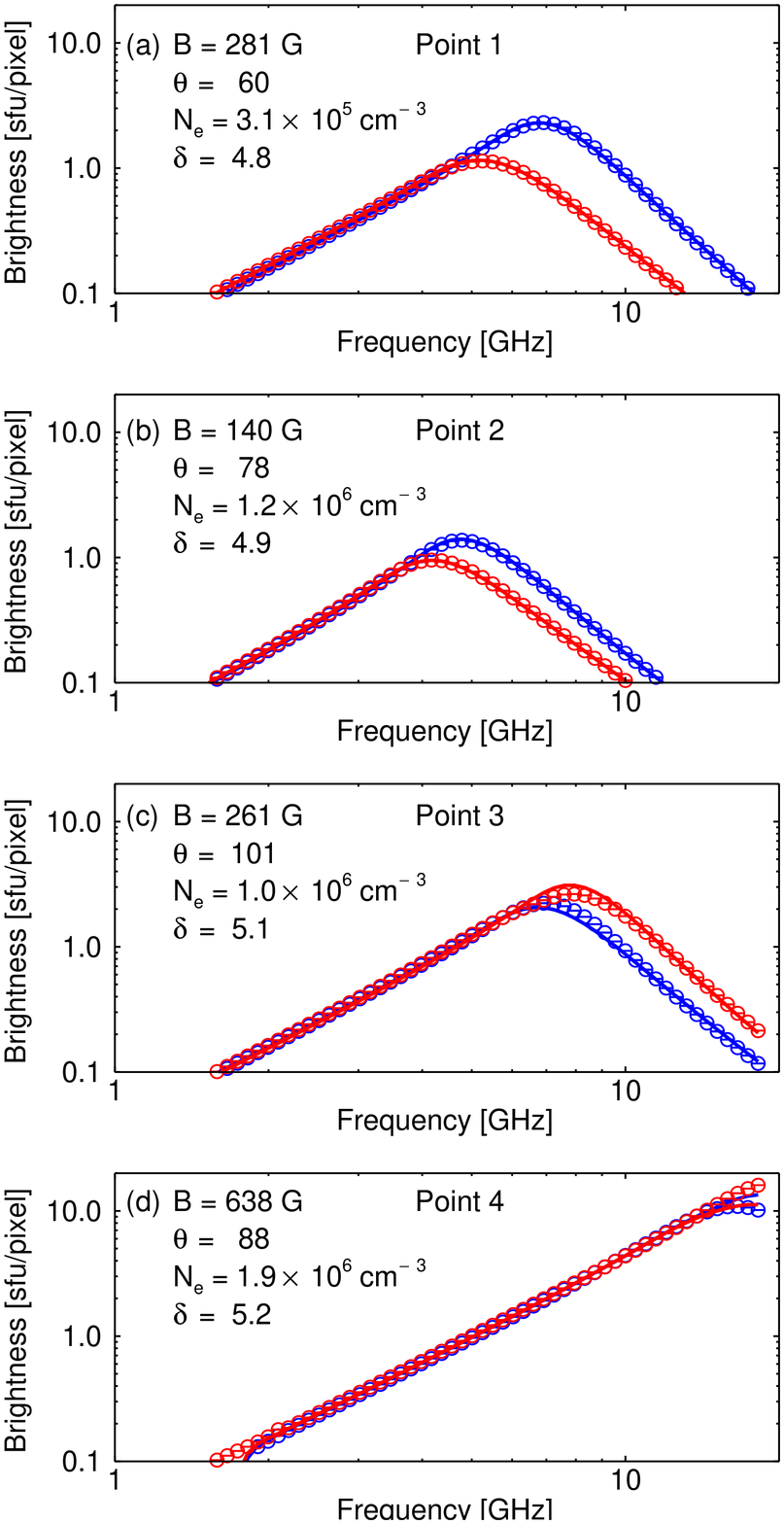}
            \includegraphics[width=0.4\textwidth,clip=,
            bb=150 -200 450 300]{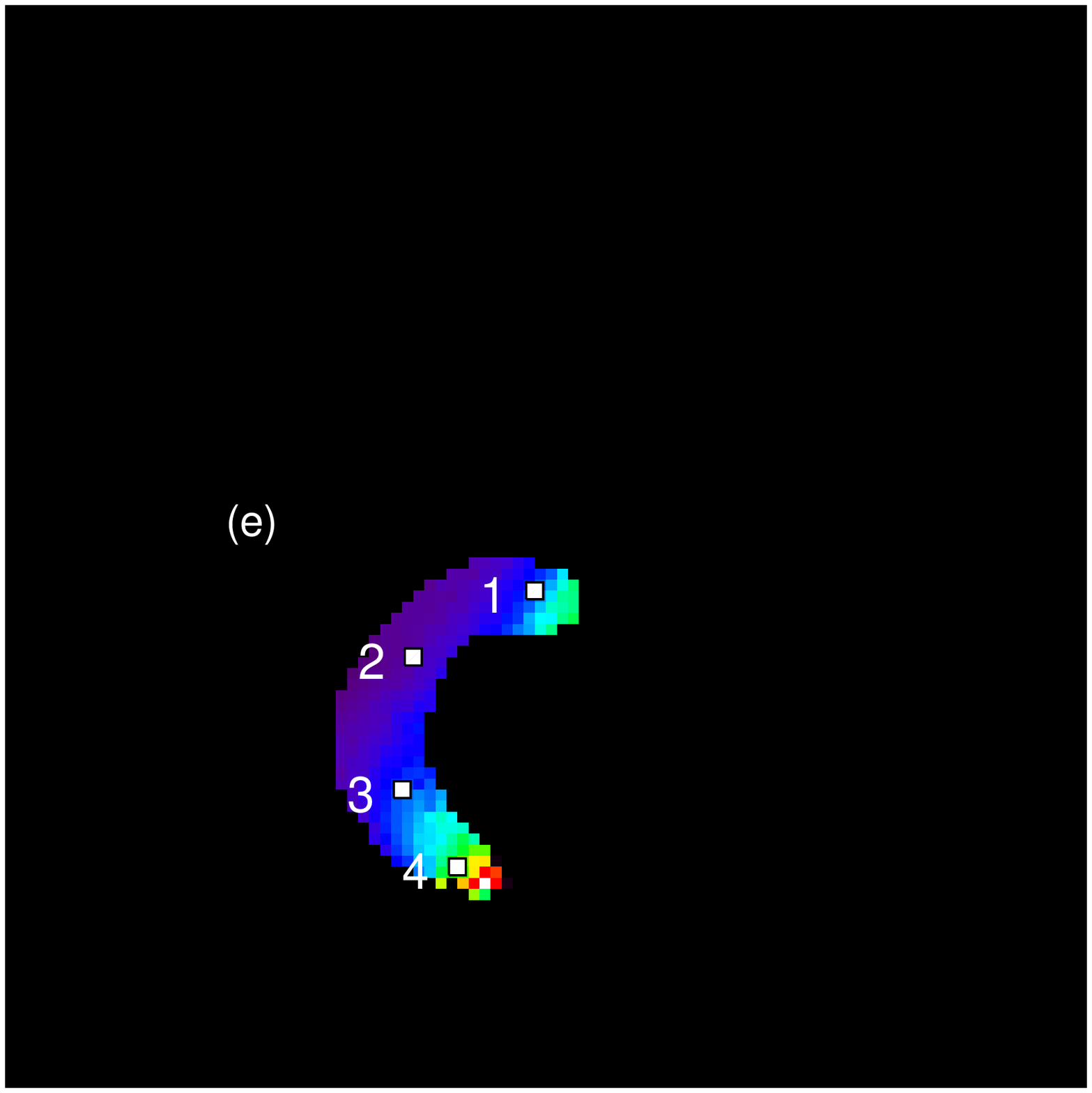} }
\caption{($a$-$d$) Overlay of spectra (symbols) from the unfolded data cube and the fits (lines) for the same four selected pixels as in Fig.~\ref{Raw_spec}.  Red corresponds to RCP, while blue is LCP. ($e$) An example 2D parameter map (total $B$) derived from the fits to the unfolded data cube, with the locations of the four points in $a$-$d$ indicated by white squares.}
\label{Unfolded_spec}
\end{figure}

\begin{figure}    
\centerline{\includegraphics[width=0.5\textwidth,clip=]{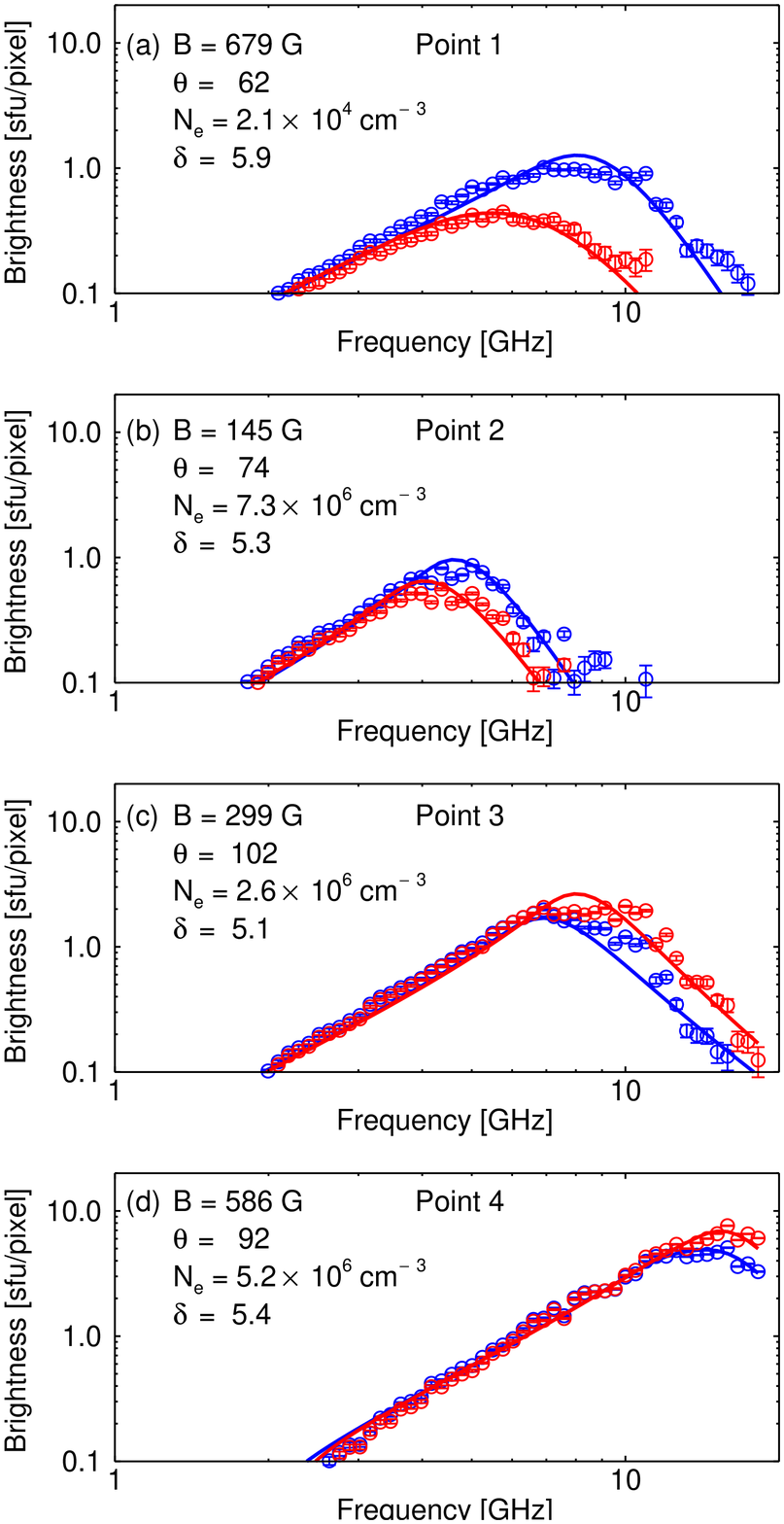}
            \includegraphics[width=0.4\textwidth,clip=,
            bb=150 -200 450 300]{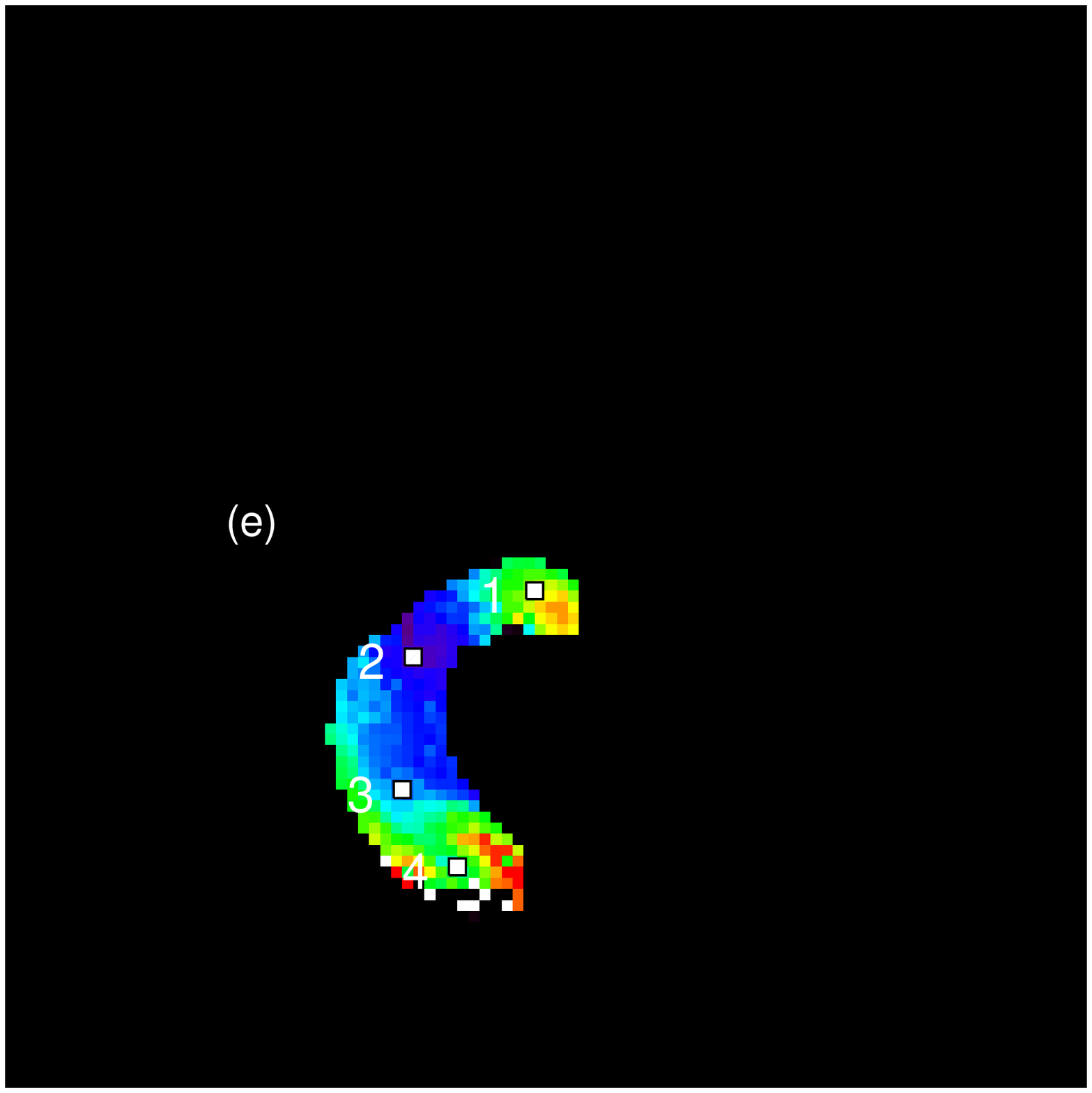} }
\caption{($a$-$d$) Overlay of spectra (symbols) from the folded data cube and the fits (lines) for the same four selected pixels as in Fig.~\ref{Raw_spec}.  Red corresponds to RCP, while blue is LCP. ($e$) The 2D parameter map of total $B$ derived from the fits to the folded data cube, with the locations of the four points in $a$-$d$ indicated by white squares.}
\label{Folded_spec}
\end{figure}

For illustration, Figure~\ref{Unfolded_spec}$a$-$d$ shows the same four "observed" dual-polarization spectra from the unfolded data cube as in Fig.~\ref{Raw_spec}$a$-$d$, with the minimum $\chi^2_\nu$ fits overlaid, while Figure~\ref{Folded_spec}$a$-$d$ shows the same for the folded data cube. Four of the parameters of each fit are shown in each panel, magnetic field strength $B$, angle of $B$ to the line of sight $\theta$, number density of nonthermal electrons $N_e$, and electron powerlaw index $\delta$.  Because the fits are done at every pixel in the 2D image cube, each parameter can be displayed as a 2D parameter map, as shown for the magnetic field strength parameter $B$ in Figs.~\ref{Unfolded_spec}$e$ and \ref{Folded_spec}$e$.  Note that although the fits in Fig.~\ref{Unfolded_spec}$a$-$d$ look excellent, they are nevertheless an approximation since the spectral points are based on radiative transfer through an inhomogeneous 3D model while the fits are done assuming a homogeneous source, albeit an independent one at each pixel.

The homogeneous source fits to the folded spectra in Fig.~\ref{Folded_spec}$a$-$d$ also adequately fit the points in each spectrum, but the fitted parameters listed in each panel differ from those of the unfolded data cube from which they are derived.  Table~\ref{Table-1} shows the comparison of parameters for the four points sampled in the figures.  Column (4) is the percent difference in folded (3) and unfolded (2) $B$, while column (7) is the difference in folded (6) and unfolded (5) powerlaw index $\delta$, and column (10) is the ratio of folded (9) and unfolded (8) energetic particle density $N_e$. The fitted parameters for Points 2-4 agree quite well--the magnetic field strengths agree to within 15\%, the powerlaw indexes agree to within 0.5, and the densities agree to well within an order of magnitude; the latter is remarkable because in addition to differences in spectra themselves, variations of two other parameters, $E_{\rm min}$ and $\delta$, recovered differently for the folded and unfolded models, also affect the value of $N_e$.  However, the agreement is worse for Point 1, due to the finite resolution of the folded images for the compact source at the northern footpoint of the loop (Fig.~\ref{Data_Cube}), which results in a considerable shift of the spectral peak in the folded compared with unfolded spectra, apparent from Fig.~\ref{Raw_spec}$a$.  We discuss these results further in the next section, as well as compare both to the parameters actually used in the model.

\begin{table}
\caption{ Comparison of fit parameters for unfolded (Fig.~\ref{Unfolded_spec}$a$-$d$) and folded (Fig.~\ref{Folded_spec}$a$-$d$) spectra.
}
\label{Table-1}
\begin{tabular}{cccccccccc}     
  \hline                   
(1) & (2) & (3) & (4) & (5) & (6) & (7) & (8) & (9) & (10) \\
Point & $B_{\rm unf}$ & $B_{\rm f}$ & $\Delta B_{\rm \%}$ & $\delta_{\rm unf}$ & $\delta_{\rm f}$ & $\Delta\delta$ & $N_{e,{\rm unf}}$ & $N_{e,{\rm f}}$ & $N_{e,{\rm rat}}$ \\
  \hline
1 & 281 & 679 & 141 & 4.8 & 5.9 & +1.1 & $3.1\times10^5$ & $2.1\times10^4$ & 0.07 \\
2 & 140 & 145 & ~3.5 & 4.9 & 5.3 & +0.4 & $1.2\times10^6$ & $7.3\times10^6$ & 6.1 \\
3 & 261 & 299 & 14.6 & 5.1 & 5.1 & ~0.0 & $1.0\times10^6$ & $2.6\times10^6$ & 2.6 \\
4 & 638 & 586 & -8.2 & 5.2 & 5.4 & +0.2 & $1.9\times10^6$ & $5.2\times10^6$ & 2.7 \\
  \hline
\end{tabular}
\end{table}

\section{Results\label{results}}
\begin{figure}    
\centerline{\includegraphics[width=0.5\textwidth,clip=]{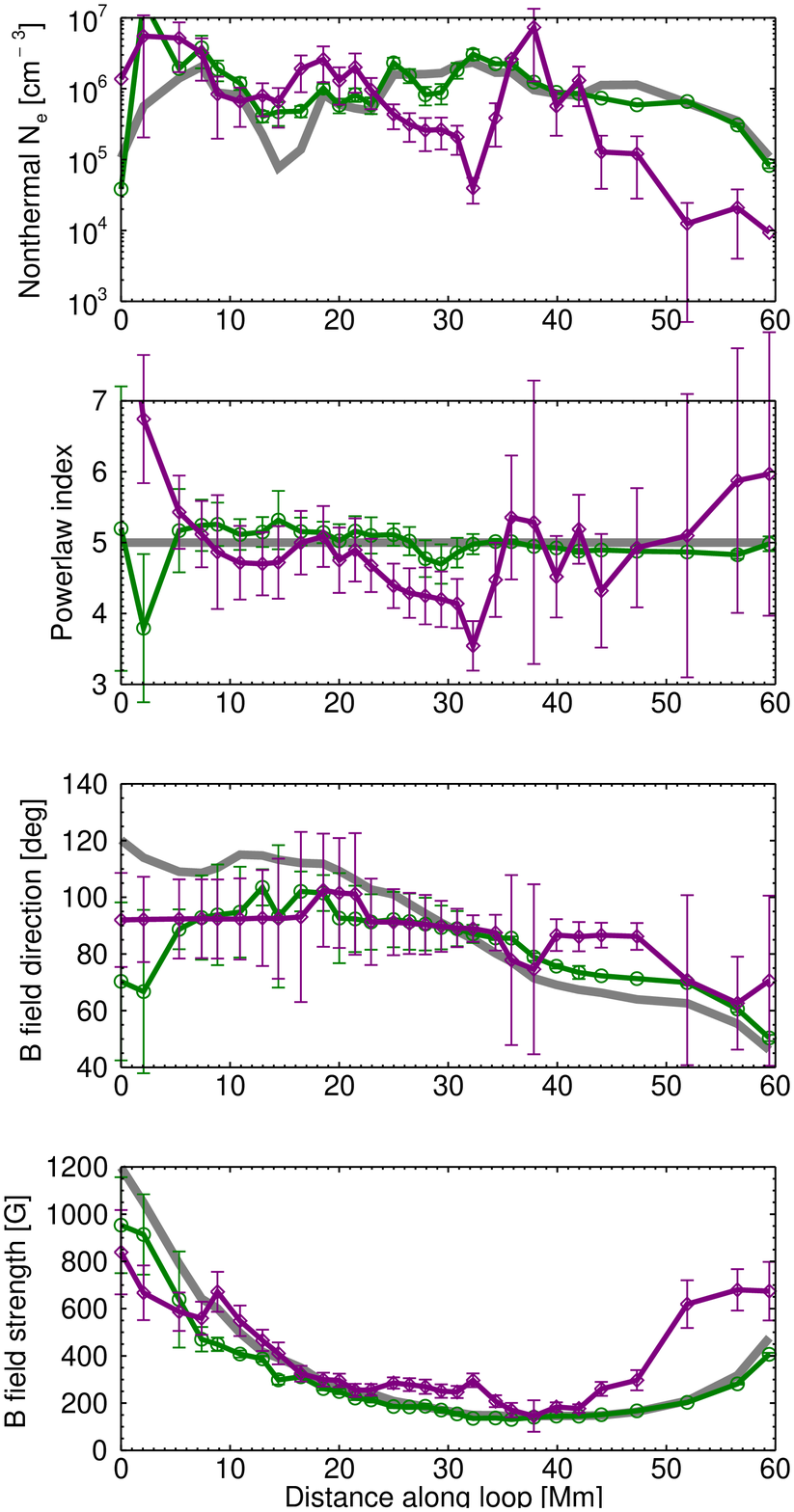}
            \includegraphics[width=0.4\textwidth,clip=,
            bb=150 -200 450 300]{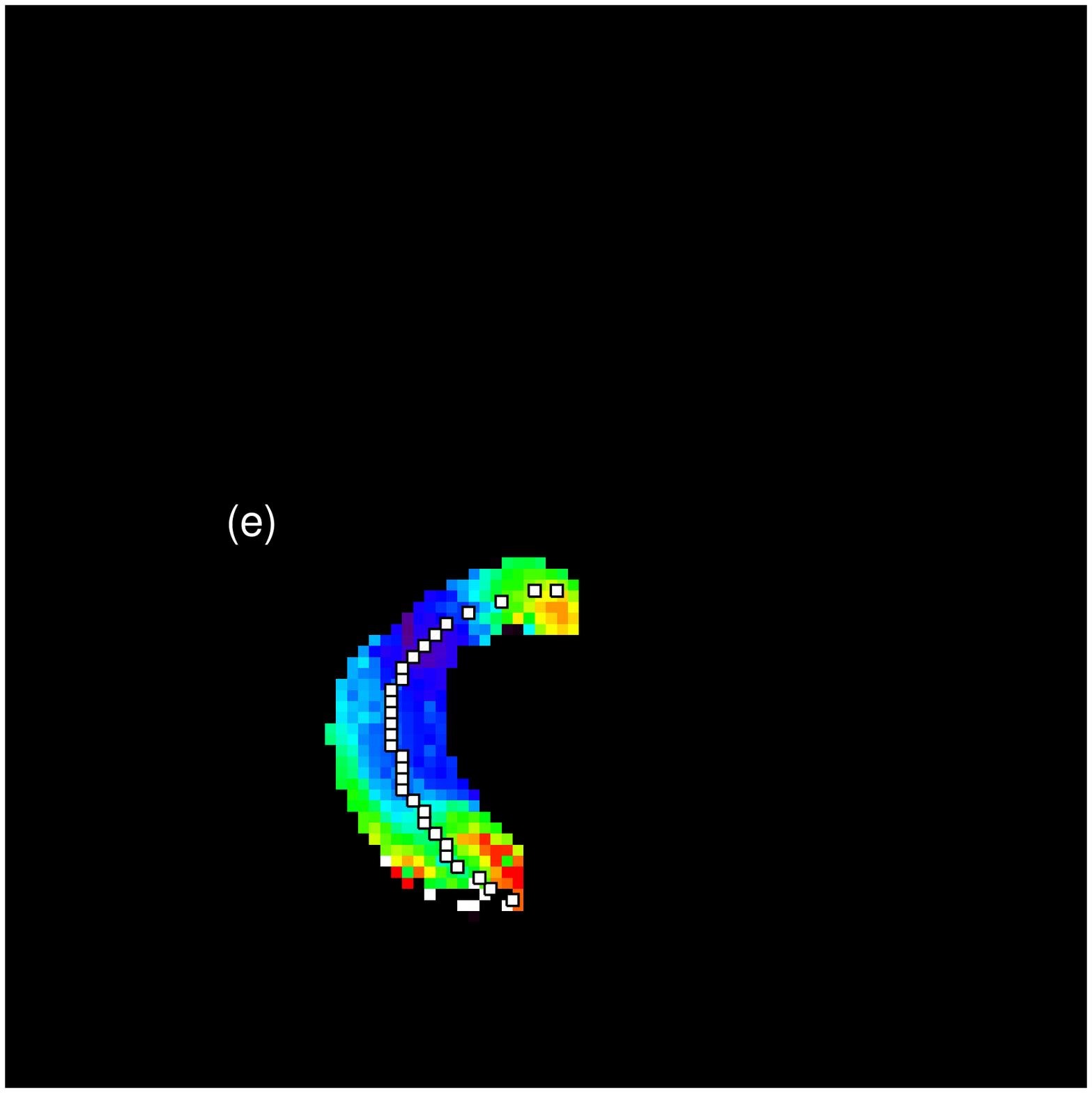} }
\caption{Comparison of parameters derived from the model (thick gray line), fits to the unfolded data (green line with open circles), and fits to the folded data (purple line with open diamonds) for positions (Mm) along the loop shown by the white squares in $e$.  The zero position corresponds to the southern footpoint of the model loop.  The error bars shown are proportional to the formal errors of the corresponding spectral fits. ($a$) Model and fit parameters for energetic electron density $N_e$. ($b$) Model and fit parameters for powerlaw index $\delta$. ($c$) Model and fit parameters for angle $\theta$ of $B$ relative to the line of sight. ($d$) Model and fit parameters for magnetic field strength $B$.  ($e$) Same 2D parameter map as in Figs.~\ref{Unfolded_spec}$e$ and \ref{Folded_spec}$e$, but now with the 29 points along the spine of the loop, indicated by white squares, whose parameters are plotted in $a$-$d$.}
\label{Profile}
\end{figure}

Comparing the $B$ parameter maps of Figs.~\ref{Unfolded_spec}$e$ and \ref{Folded_spec}$e$, it is clear that the best agreement is along the central spine of the loop, with worse agreement at the edges where the effects of finite resolution of the folded images is most pronounced.  To investigate the potential for obtaining spatially-resolved parameters along a flaring loop, we show in Figure~\ref{Profile} the run of parameters along the spine of the loop, at the locations of the points shown in Fig.~\ref{Profile}$e$.  In agreement with the results of \inlinecite{Fl_etal_2009}, we find that the fit parameters from the unfolded data cube (green lines) in Fig.~\ref{Profile}$a-d$, representing data from an "ideal" radio heliograph with infinite spatial resolution, agree very well with the model (gray lines) at nearly every point.  What is new in this work, however, are the purple lines representing the parameters derived from the folded data cube with its finite resolution.  Even in this case, the magnetic field strength (Fig.~\ref{Profile}$d$) agrees well qualitatively over the entire length of the loop, and even quantitatively over about 2/3 of the loop (the region from 6-47 Mm in the figure).  Likewise, the other parameters also agree tolerably well except at each end of the loop where distortions in the spectra caused by the finite resolution have the largest effect.  Note that the error bars derived from the goodness of the spectra fits increase near the ends of the loop, and thus give a reasonable indication of where the parameters are less well determined.

\begin{figure}    
\centerline{\includegraphics[width=0.5\textwidth,clip=,bb=50 30 590 400]{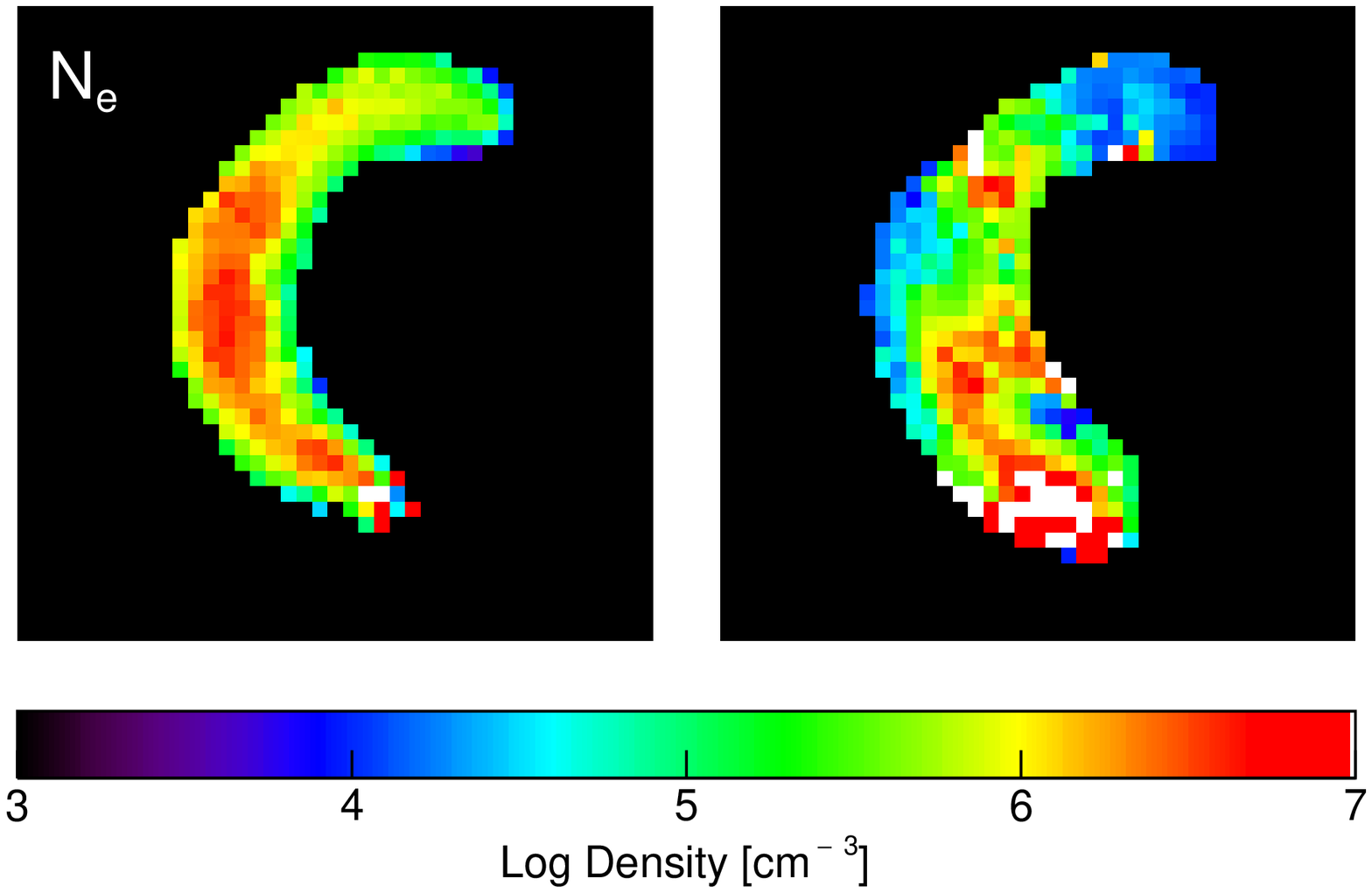}
            \includegraphics[width=0.5\textwidth,clip=,bb=50 30 590 400]{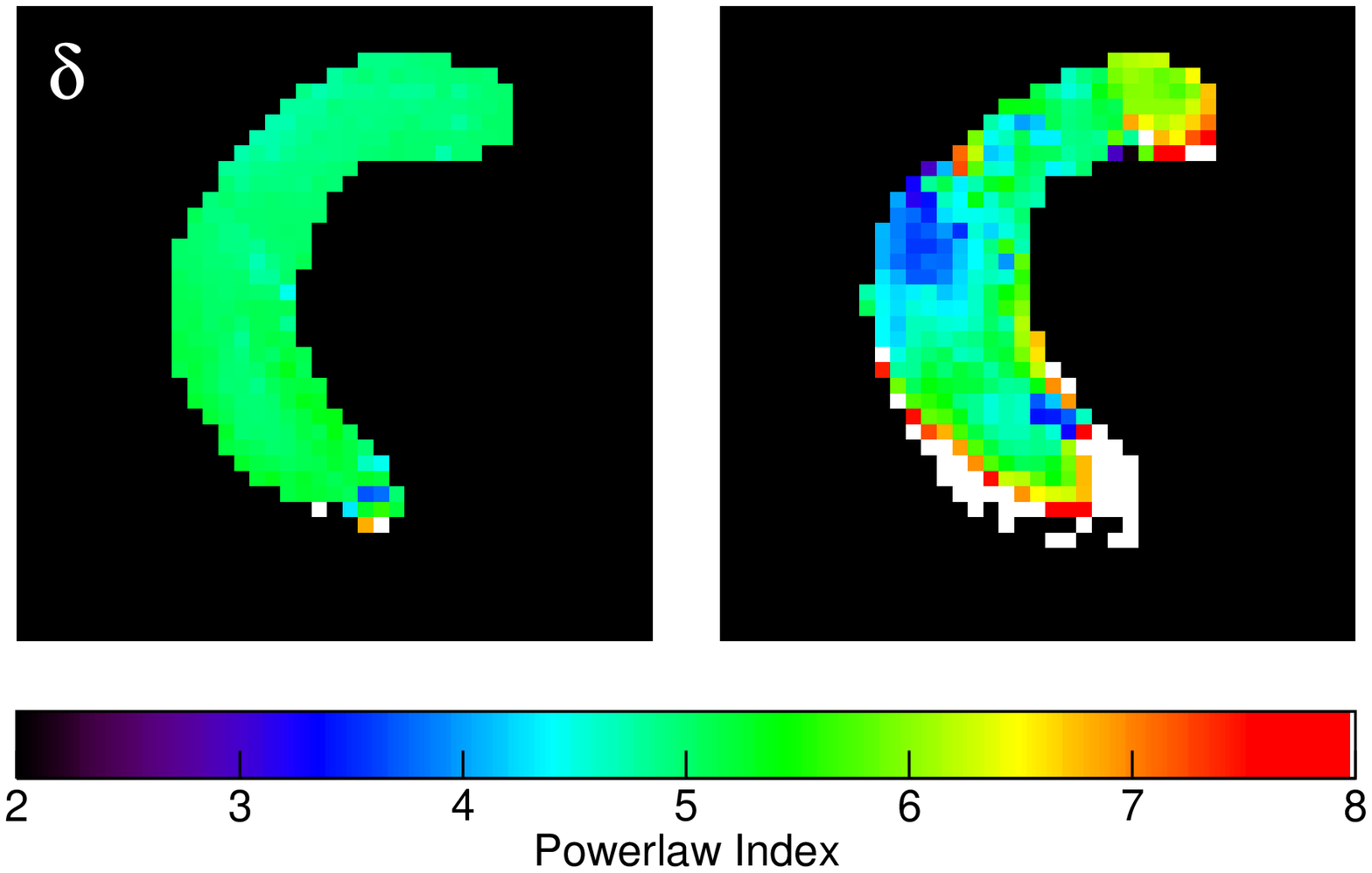} }
\centerline{\includegraphics[width=0.5\textwidth,clip=,bb=50 30 590 400]{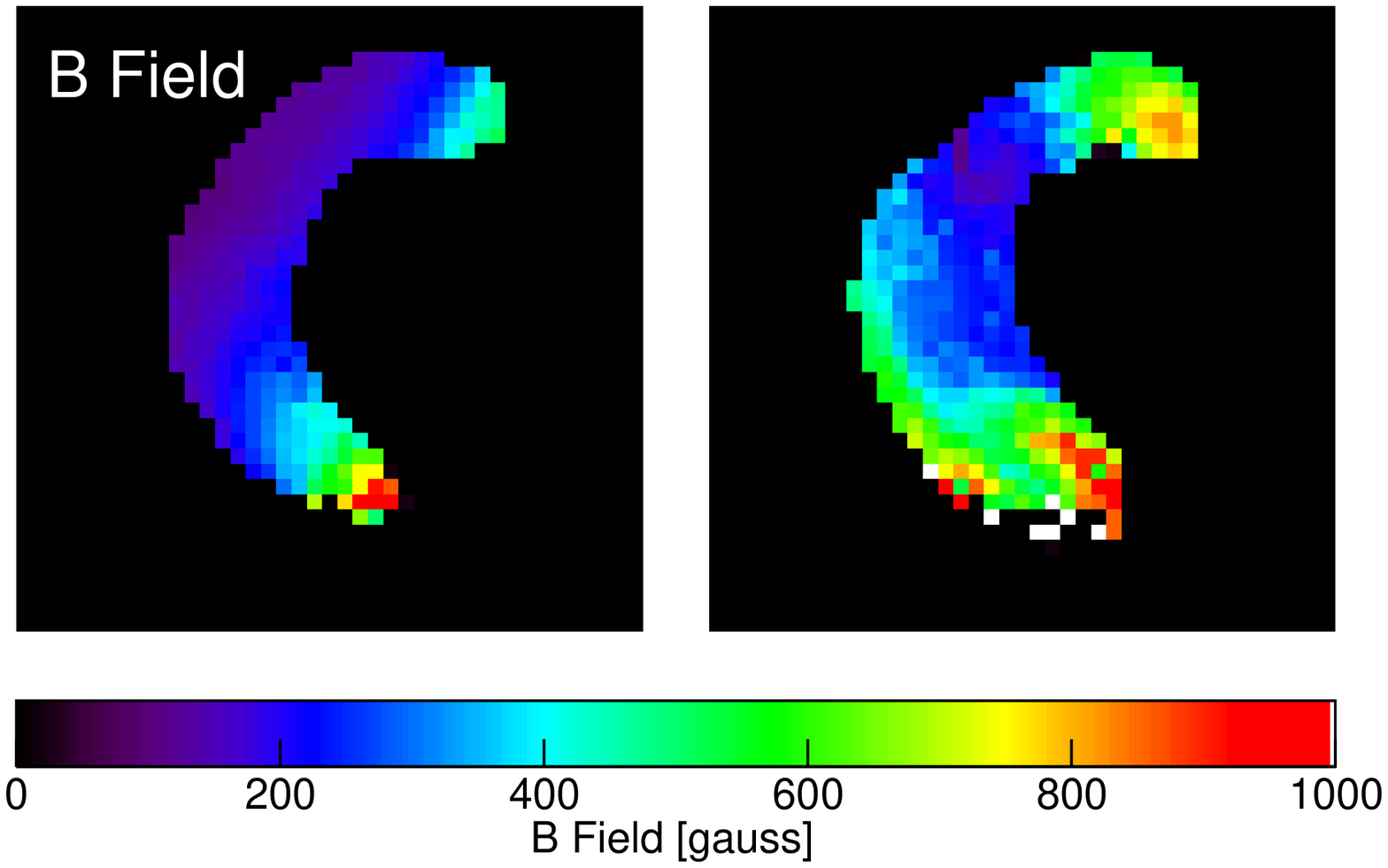}
            \includegraphics[width=0.5\textwidth,clip=,bb=50 30 590 400]{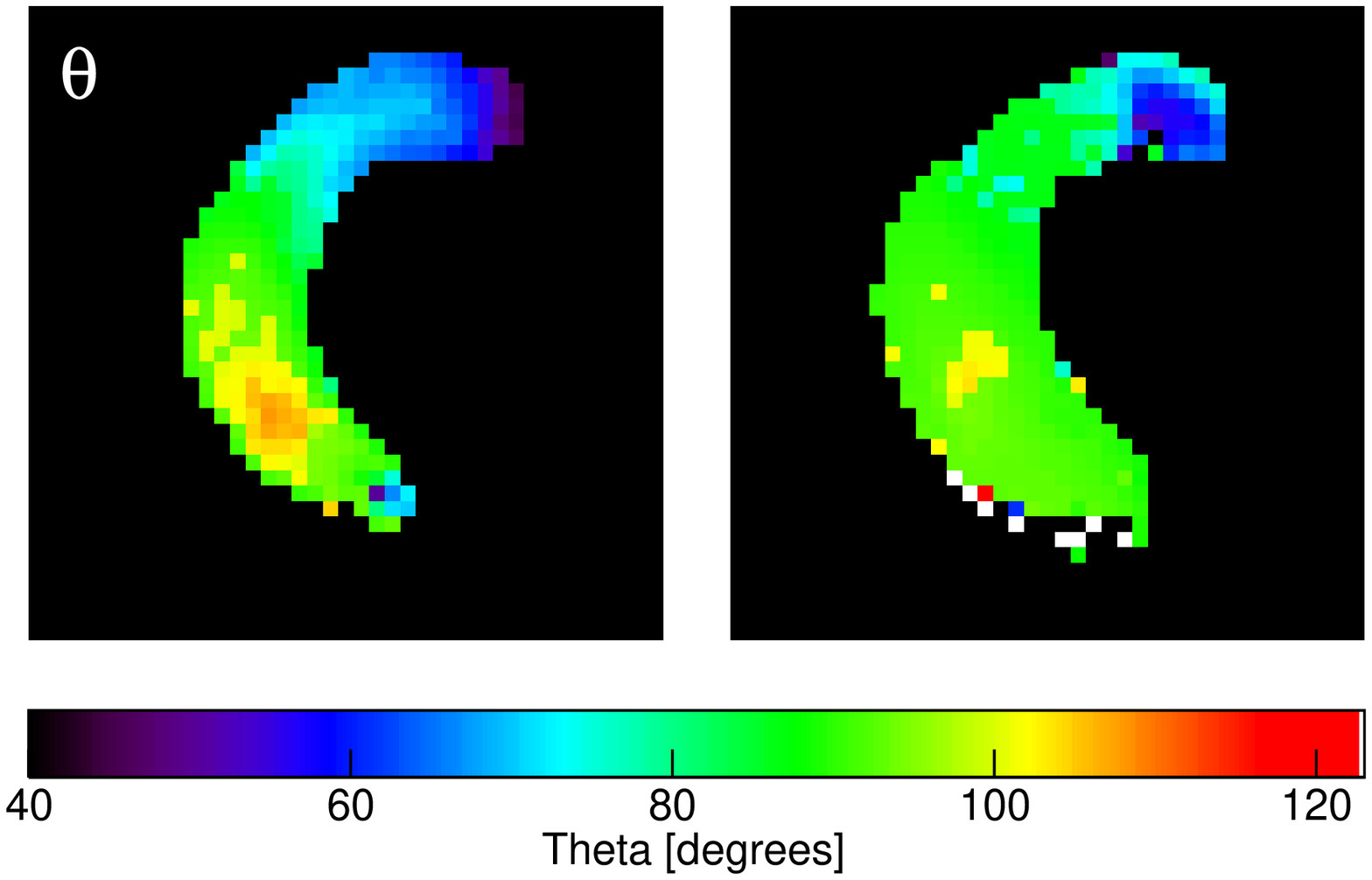} }
\caption{Comparison of 2D parameter map pairs for the unfolded (left) and folded (right) data cubes in each of four parameters.  Log density of nonthermal electrons $N_e$ is shown at upper left, powerlaw index $\delta$ at upper right, $B$ field at lower left, and angle $\theta$ of $B$ to the line of sight at lower right.  Edge effects due to finite resolution are evidence in the folded parameter maps, which are largely absent in the unfolded maps.}
\label{Parameters}
\end{figure}

Another useful comparison is of the 2D parameter maps themselves, which are shown in pairs (unfolded on the left and folded on the right of each pair) in Figure~\ref{Parameters}.  Below each pair is a color bar showing the range of the depicted parameter.  Arguably the best-determined parameters, at least along the spine of the loop, are the magnetic parameters $B$ and $\theta$, due to the fact that the key spectral features that determine these, the location of the peak frequency and the polarization, respectively, are robustly measured in the folded spectra.  Although $N_e$ and $\delta$ vary more erratically along the loop, their general values still compare well with those of the folded maps and the model, and represent a huge advance over what has been possible with existing radio instruments.

\section{Discussion and Conclusions\label{conclusions}}

\begin{figure}    
\centerline{\includegraphics[width=0.9\textwidth,clip=]{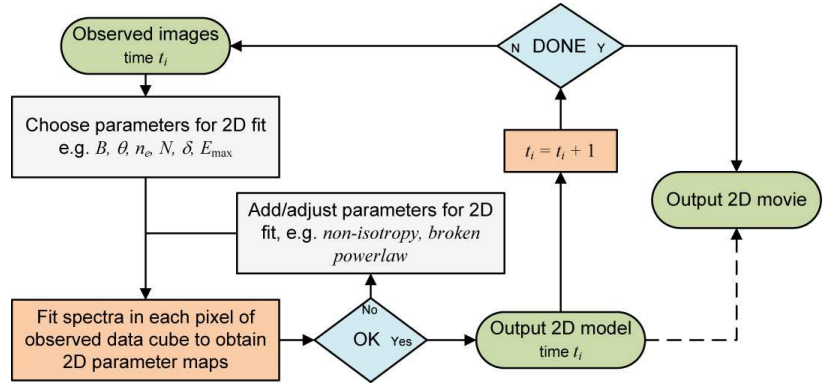}}
\caption{Block diagram showing the steps performed to create 2D parameter map movies from the data cube obtained from an actual instrument (the folded data cube).}
\label{Alg_2}
\end{figure}

We note that the result shown in Fig.~\ref{Parameters} is for a single instant in time, and in the case of EOVSA can be repeated for independently measured spectra once per second to yield a powerful new tool for following the dynamically changing particle and field parameters in flaring loops. We conclude that with the advent of new, broadband microwave imaging instruments, the technique of microwave imaging spectropolarimetry will soon become a viable and important means of obtaining dynamic, 2D parameter maps of flares.  Figure~\ref{Alg_2} shows the procedure one might use to obtain movies of such 2D parameter maps, which requires no 3D modeling and employs only measured radio data.  One simply starts with the observed, multifrequency images, and fits the polarized spectra in each pixel of the images using a specified set of parameters as we have shown above.  If the $\chi_\nu^2$ of the fits is generally not close to unity, it will suggest that some important physics is missing, and the addition of such parameters as pitch-angle non-isotropy and/or alteration of the electron energy distribution is warranted, which can be added based on physically motivated trial-and-error.  Once the fits are acceptable for a given time sample, the procedure is repeated for the next time step until the dynamic maps are obtained.

Although such a procedure is possible, and is a huge advance over what has been possible before, our results above have shown that finite resolution, finite dynamic range, and systematic effects from image reconstruction all play a role in distorting the measured microwave spectra.  These effects can be reduced by improving the resolution and imaging capability of the instrument, which is an important motivation for constructing the high-performance Frequency Agile Solar Radiotelescope (FASR, e.g. \opencite{Gary_Keller_2004}).  Even with FASR, however, the measured spectra would still represent emission from a range of parameters along the line of sight while the parameter fitting we have described is done using the assumption of a homogeneous source.  Meanwhile, a great deal of useful information on magnetic field, thermal plasma and energetic particles is available from spacecraft and groundbased instruments, but has gone unused in the above scenario.

\begin{figure}    
\centerline{\includegraphics[width=0.9\textwidth,clip=]{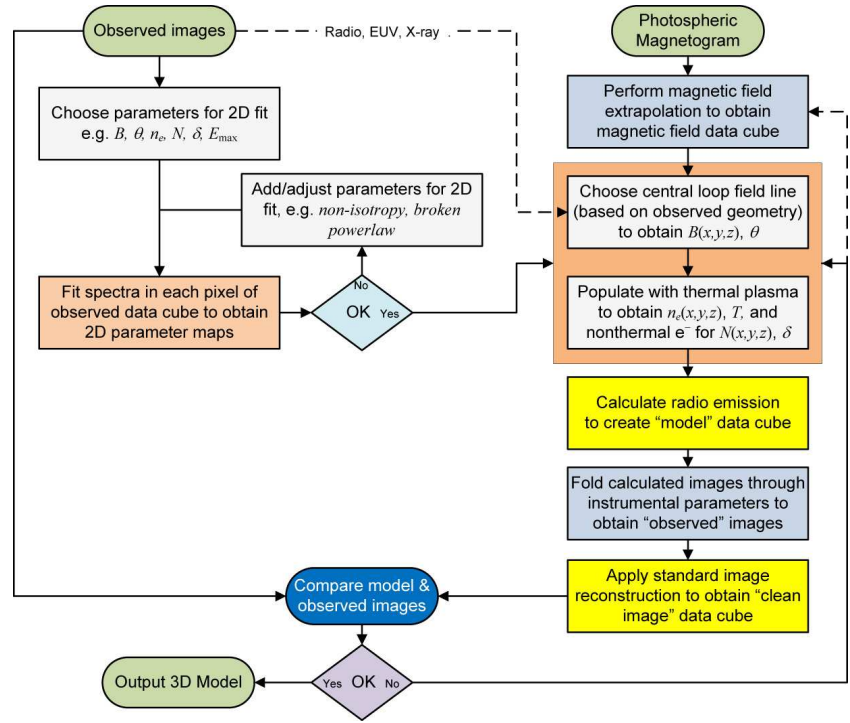}}
\caption{Block diagram that incorporates the 2D parameter maps obtained by the procedure of Fig.~\ref{Alg_2} into a larger, integrated scheme of 3D modeling that also uses all available data from other wavelength regimes.}
\label{Alg_3}
\end{figure}

Figure~\ref{Alg_3} integrates the 2D information from Fig.~\ref{Alg_2} with elements of the 3D modeling from Fig.~\ref{Alg_1}, and also incorporates all available data from other wavelengths.  In this scheme, the 2D parameter maps merely provide a guide to 3D modeling. The 3D geometry starts from a model of the magnetic field, which is shown in Fig.~\ref{Alg_3} as a magnetic field extrapolation, but could be obtained by some other means such as MHD numerical modeling.  Within this geometry one chooses the relevant field regions involved in flaring, utilizing morphological information from radio, EUV, X-rays, or other data.  One then populates the magnetic model with a suitable population of plasma and energetic particles, and it is here that the 2D parameter maps provide guidance, along with parameters obtained similarly from other wavelength regimes.  The key step is to then use the populated 3D model to calculate emission (shown as radio emission in the figure,  but EUV, X-ray and other emissions can also be calculated from the model) and fold it through the relevant instrument(s) making the observations.  The ultimate comparison is then done with the observed images and the folded model images.

This approach completely side-steps issues of finite resolution, finite dynamic range, image reconstruction, and even homogeneous vs. inhomogeneous sources.  If the modeled images match the observed ones, one then accepts the entire 3D model.  In the more likely case of a mismatch, however, one goes back to adjust the choice of field regions or plasma and particle populations.  It is this step that will provide the key physical insights into the flaring process.  It is here that particle energy and pitch-angle distributions predicted from various levels of wave-particle interactions may be tried, or any of a large number of other adjustments. If this procedure does not converge, it may be that one must follow the dashed line on the right, and modify the magnetic model.  It is here that key insights into the limitations and perhaps refinements of magnetic field extrapolations, MHD simulations, or even new approaches to magnetic field modeling will be stimulated. It is here that information from new magnetic field measurement techniques can be incorporated, such as chromospheric magnetograms, Zeeman measurements of infrared coronal emission, measurements using the Hanl\'e effect, and others.

This forward fitting approach is analogous to terrestrial weather modeling, and is a large and complex endeavor that will require the efforts of the entire solar community \cite{Fl_arXiv_2010}.  We anticipate that by working together we can succeed in making true progress in understanding the physics of magnetic reconnection, particle acceleration, and the other areas of research noted in Introduction.  The ability of microwave imaging spectropolarimetry to make dynamic plasma, particle and magnetic field measurements, which we have demonstrated here, makes it an important tool in this endeavor.


\begin{acks}
This work was supported in part by NSF grants
AGS-0961867, AST-0908344, AGS-1250374 and NASA grants NNX10AF27G and NNX11AB49G to New Jersey
Institute of Technology.  This work also benefited from workshop support from the International Space Science Institute (ISSI). 
%
%
%
\end{acks}

%
%
 \bibliographystyle{spr-mp-sola}
\bibliography{fleishman,xray_refs,ms_bib,magnetography,radio_synth_imaging}
%
%
%

\end{article}
\end{document}